\documentclass[iop,twocolappendix]{emulateapj}

\shorttitle{Serpens X-1}
\shortauthors{Chiang et al.}

\begin{document}
\title{A test of the nature of the Fe K Line in the neutron star low-mass X-ray binary Serpens X-1}
\author{Chia-Ying Chiang\altaffilmark{1}, Edward M. Cackett\altaffilmark{1}, Jon M. Miller\altaffilmark{2}, 
Didier Barret\altaffilmark{3,4}, Andy C. Fabian\altaffilmark{5}, Antonino D'A\`{i}\altaffilmark{6},\\ Michael L. Parker\altaffilmark{5},
Sudip Bhattacharyya\altaffilmark{7}, Luciano Burderi\altaffilmark{8}, Tiziana Di Salvo\altaffilmark{9}, 
Elise Egron\altaffilmark{10},\\ Jeroen Homan\altaffilmark{11}, Rosario Iaria\altaffilmark{9}, 
Dacheng Lin\altaffilmark{12}, and M. Coleman Miller\altaffilmark{13}}
\affil{$^1$Department of Physics and Astronomy, Wayne State University,
    666 W. Hancock, Detroit, MI 48202, USA}
\affil{$^2$Department of Astronomy, The University of Michigan, 500 Church Street, Ann Arbor, MI48109-1046, USA}
\affil{$^3$Universite de Toulouse, UPS-OMP, Toulouse, France}
\affil{$^4$CNRS, Institut de Recherche en Astrophysique et Planetologie, 9 Av. colonel Roche, BP 44346, F-31028 Toulouse cedex 4, France}
\affil{$^5$Institute of Astronomy, University of Cambridge, Madingley Road, Cambridge CB3 0HA, UK}
\affil{$^6$INAF-Istituto di Astrofisica Spaziale e Fisica Cosmica di Palermo, via U. La Malfa 153, 90146 Palermo, Italy}
\affil{$^7$Department of Astronomy and Astrophysics, Tata Institute of Fundamental Research, Mumbai 400005, India}
\affil{$^8$Dipartimento di Fisica, Universit{\`a} degli Studi di Cagliari, SP Monserrato-Sestu, KM 0.7, I-09042 Monserrato, Italy}
\affil{$^9$Dipartimento di Fisica e Chimica, Universit\'{a} di Palermo, via Archirafi 36, I-90123 Palermo, Italy}
\affil{$^{10}$INAF-Osservatorio Astronomico di Cagliari, via della Scienza 5, 09047 Selargius (CA), Italy}
\affil{$^{11}$MIT Kavli Institute for Astrophysics and Space Research, 77 Massachusetts Avenue 37-582D, Cambridge, MA 02139, USA}
\affil{$^{12}$Space Science Center, University of New Hampshire, Durham, NH 03824, USA}
\affil{$^{13}$Department of Astronomy and Joint Space-Science Institute, University of Maryland, College Park, MD 20742-2421 USA}

\begin{abstract}
Broad Fe K emission lines have been widely observed in the X-ray spectra 
of black hole systems, and in neutron star systems as well. The intrinsically 
narrow Fe K fluorescent line is generally believed to be part of the reflection 
spectrum originating in an illuminated accretion disk, and broadened by 
strong relativistic effects. However, the nature of the lines in neutron star 
LMXBs has been under debate.  We therefore obtained the longest, 
high-resolution X-ray spectrum of a neutron star LMXB to date with a 300 ks 
\emph{Chandra} HETGS observation of Serpens X-1.  The observation was 
taken under the ``continuous clocking" mode and thus free of photon pile-up 
effects. We carry out a systematic analysis and find that the blurred reflection 
model fits the Fe line of Serpens X-1 significantly better than a broad Gaussian 
component does, implying  that the relativistic reflection scenario is much preferred. 
\emph{Chandra} HETGS also provides highest spectral resolution view of the 
Fe~K region and we find no strong evidence for additional narrow lines.

\end{abstract}


\section{Introduction}

Broad iron emission lines have been widely discovered in Active Galactic Nuclei 
(AGN; \citealt{Tanaka95,Fabian09,Brenneman11}), and low-mass X-ray binaries 
(LMXBs) including stellar-mass black holes \citep{DZ99,Miller02,Miller07,Reis09}
and neutron stars (NS; \citealt{Asai00,Barret00,Oosterbroek01,DS05,Bhattacharyya07,Iaria07,Cackett08,Cackett09,Cackett10,Sanna13,DiSalvo15,Pintore15}). 
An accretion disk is believed to orbit the central object, and a hard X-ray source, 
either a powerlaw continuum or a blackbody component (potentially the 
``boundary layer"), emits hard X-rays that illuminate the accretion disk. Atomic 
transitions take place after the high-energy photons are absorbed, resulting in a 
reflection spectrum including several narrow emission lines and a broad feature 
peaked around 20-30 keV which is known as ``Compton hump" 
\citep{LW88,GF91,Ross93,Matt93,Ross05,GK10,Ballantyne12}. 

The Fe K$\alpha$ fluorescent line is the most prominent feature in the reflection 
spectrum. When appearing in the X-ray spectra of AGN and LMXBs, the 
intrinsically narrow Fe lines sometimes show broad, asymmetric profiles which 
are generally believed to be shaped by a series of relativistic effects induced from 
strong gravitational fields \citep{Fabian89,Fabian00}. As relativistic effects are stronger 
in the area closer to the compact object, the line profile is sensitive to the inner 
radius of the accretion disk. If the accretion disk extends to the innermost stable 
circular orbit (ISCO), one can, under certain assumptions, obtain an estimate of 
the black hole spin by measuring the inner radius of the accretion disk 
\citep{Bardeen72}. The Fe K$\alpha$ line profile has been a powerful tool to 
measure black hole spin \citep[e.g.,][]{Miller02,RN03,BR06,Miller07,Reis08,Reis12}. 
The inner radius of the accretion disk in a NS system can be determined by the 
same method. The accretion disk in a NS system could be truncated by the stellar
surface or the boundary layer between the disk and the NS, if the NS is larger 
than its ISCO, or by a strong stellar magnetic field. An upper limit of the stellar 
radius of a NS can be given by measuring the inner radius of the disk, and hence 
help understand its equation of state (e.g. \citealt{Piraino00,Cackett08}). \citet{B11} 
reported more detailed calculations to show how future instruments can directly 
constrain NS equation of state models using relativistic disk lines.

In NS systems, the Fe K$\alpha$ lines are usually not as prominent (EW $\sim$ 
100 eV) as those seen in AGN and black hole binaries (BHBs) due to extra 
continuum emission from the boundary layer. It has been widely accepted that 
relativistic Fe K lines are common in AGN \citep{RN03} and BHBs \citep{Miller07}, 
though some suggest line profiles to be caused by warm absorbers \citep{IM03} or 
Comptonization \citep{LT07}. Nonetheless, whether Fe K lines in NS systems are 
relativistically broadened is still under debate. \citet{Ng10} analyzed a number of 
\emph{XMM-Newton} NS spectra and concluded that statistical evidence of 
asymmetric iron line profiles is lacking and the lines are broadened by Compton 
scattering in a disk corona \citep{Misra98,Misra99}. \citet{RW00} showed that the 
continuum source required in the Compton scattering model to produce the broad 
iron line violates the blackbody limit. Although the calculation was based on the 
AGN case, small radii of the Compton clouds are still required to maintain the high 
ionization level in NS systems, which would make gravitational effects dominant 
\citep{Fabian95}. Furthermore, the pile-up correction applied in \citet{Ng10} 
reduced the signal-to-noise ratio of the data and made it difficult to detect relativistic 
iron lines (see Figure 2 in \citealt{Miller10}). \citet{Cackett10} examined 
a large sample of \emph{Suzaku} NS spectra and found that the relativistic reflection 
scenario is much preferred. The iron line detections with \emph{Suzaku} are 
less affected by photon pile-up effects than those of \emph{XMM-Newton}, thus the 
conclusion that iron lines are asymmetric is likely more robust. A study of the effects of
pile-up in X-ray CCD detectors by \citet{Miller10} showed that while pile-up can distort
the Fe K line profiles, it tends to artificially narrow them (in contrast with \citealt{Ng10}).

Detections made with spectrometers that suffer no photon pile-up effects become 
important to determine the iron line profiles. The re-analyses of archival 
\emph{BeppoSAX} data of 4U 1705-44 implied the existence of asymmetric Fe K lines 
\citep{Piraino07,Lin10,Cackett12,Egron13}. \emph{NuSTAR} sees an asymmetric line
in Serpens X-1 \citep{Miller13} with line 
properties consistent with \emph{Suzaku} measurements from \citet{Cackett08,Cackett10,Cackett12}.
The pile-up free detectors \emph{BeppoSAX} and \emph{NuSTAR} both reach the 
conclusion of relativistic iron lines. These instruments are, however, not capable of 
detecting possible narrow line components on top of the broad Fe K lines. 
\emph{Chandra} HETGS is so far the only instrument that offers a pile-up free observation 
at high spectral resolution (approximate $\Delta$E$\sim$30 eV at 6 keV).

The neutron star LXMB Serpens X-1 was discovered in 1965 \citep{Friedman67}. Being 
a persistent, bright X-ray source, Serpens X-1 has been observed with major X-ray 
missions, including \emph{Einstein} \citep{Vrtilek86}, \emph{ASCA} \citep{Asai00}, 
\emph{EXOSAT} \citep{Seon02}, \emph{BeppoSAX} \citep{Oosterbroek01}, 
\emph{INTEGRAL} \citep{Masetti04}, \emph{XMM-Newton} \citep{Bhattacharyya07}, 
\emph{Suzaku} \citep{Cackett08,Cackett10} and recently with \emph{NuSTAR} \citep{Miller13}. 
It was also detected in optical \citep{Hynes04} and radio \citep{Migliari04} band. 
Relativistic Fe K lines have been reported several times in previous \emph{Suzaku}, 
\emph{XMM-Newton} and \emph{NuSTAR} observations 
\citep{Bhattacharyya07,Cackett08,Cackett10,Miller13}. In this paper we study the latest high-resolution 
\emph{Chandra} HETGS observation, which is the longest \emph{Chandra} grating 
observation of a neutron star LMXB to date. We present detailed data analysis and 
results in the following sections. The Galactic absorption column $N_{\rm H}$ is 
assumed to be $4.4\times10^{21}$ cm$^{-2}$\citep{DL90} with ``wilm" abundances 
\citep{Wilms00} throughout all our analyses, which were done using the XSPEC 12.8.2 package \citep{Arnaud96}. 
All errors quoted in the paper are given at the 90 
per cent confidence level.

\section{Data Reduction}\label{sec:datared}

Serpens X-1 was observed with the \emph{Chandra} High Energy Transmission 
Grating Spectrometer (HETGS) during 2014 June $27-29$ and 2014 August $25-26$ 
(Obs. ID: 16208, 16209), totaling a good exposure of $\sim$ 300 ks. The observation 
was taken using the ``continuous clocking (CC)" mode. 
We reduced the data following the standard procedures using the latest CIAO V4.6 
software package. The CC mode provides 2.85 ms time resolution, the observation is 
hence clear of photon pile-up effects and has negligible backgrounds. In this work we 
concentrate on the HEG (high energy grating) data which covers the Fe line 
energy band.  We found the HEG +1 and $-1$ spectra to differ at the $5-10$\% level, especially
around the area of a chip gap in the +1 spectrum.  Thus, we tested a number 
of methods to improve the +1 spectrum (see Appendix). However, the discrepancy
cannot be completely eliminated, and thus we only use the HEG $-1$ spectrum in this work.

The spectra of each observation from June and August 2014 are similar and we 
combined them to form a long spectrum. All spectral bins from 2-8 keV have more than
30 counts per bin and thus no rebinning is required. 
We found a number of
wiggles in the spectrum below 2 keV that could not be modeled. Their locations matched those
where there are significant sharp changes in the effective area.  We therefore use the
$2.0-8.0$ keV energy band in the following analysis. A restricted energy band is often 
 used for the data taken under the CC mode to avoid artificial instrumental
artifacts \citep[e.g.,][]{Cackettcc,miller11,miller12,Degenaar14}.

\begin{figure}
\begin{center}
\includegraphics[scale=0.5]{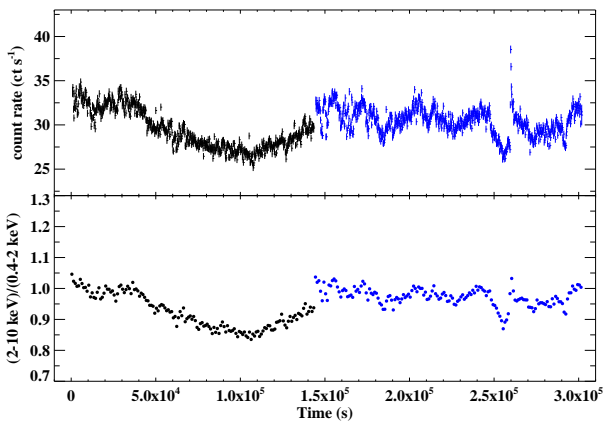}
\caption{The upper panel shows the 0.4-10 keV light curve of Serpens X-1. 
The light curve was extracted from the first order HEG data.
The gap between the observations has been discarded. Black points 
stand for data extracted from the first half of the observation, and blue 
ones for the second. There is a weak type I X-ray burst
containing $\sim$ 0.5 per cent of the total counts in the second half of the observation.
The lower panel shows the hardness ratio, which is the hard (2-10 keV) light 
curve divided by the soft (0.4-2.0 keV) light curve.}
\label{lc}
\end{center}
\end{figure}

The upper panel of Fig. \ref{lc} displays the $0.4-10.0$ keV light curves of Serpens X-1 during the 
observation. It can be seen that there is a weak type I X-ray burst, which originates 
from the thermonuclear burning of matter on the NS surfaces 
\citep{Woosley76,Lamb78,Strohmayer06}, in the later observation (shown in blue data 
points). The X-ray burst lasted for a few hundreds of seconds and contributed 
$\sim 4.7\times10^{4}$ counts, which is only $\sim$ 0.5 per cent of the total 
$\sim8.9\times10^{6}$ counts of the entire observation. Since the burst is such a small 
fraction of the total counts and casts no effects on the spectrum, we did not exclude 
the data during the X-ray burst.

\begin{deluxetable*}{llrrrrrrr}
\tabletypesize{\scriptsize}
\tablecaption{Best-fitting parameters for different continuum models. Model 1a/1b includes
both DISKBB and BBODY components; Model 2a/2b composes of DISKBB and the powerlaw continuum; 
Model 3a/3b comprises of BBODY and powerlaw components. Both Model 3a and Model 3b gave most reasonable fitting parameters.
\label{continuum}
}
\startdata
\tableline\tableline
Component &Parameter & Model 1a & Model 1b &  Model 2a &  Model 2b & Model 3a & Model 3b\\
\tableline
TBABS & $N_{\rm H}$ ($10^{22}$ cm$^{-2}$)& (0.44) & (0.44) & (0.44) & (0.44) & (0.44) & (0.44)\\
DISKBB & $kT_{\rm disk}$ (keV)& $1.34^{+0.05}_{-0.03}$ & $1.42\pm0.04$ & $1.42\pm0.01$ & $1.44^{+0.01}_{-0.02}$ & \nodata  & \nodata \\
 & $N_{\rm disk}$ & $80.0^{+6.8}_{-8.3}$ & $68.3^{+6.5}_{-3.4}$ & $60.1^{+2.3}_{-3.0}$ & $61.7^{+1.4}_{-2.8}$ &\nodata & \nodata \\
BBODY & $kT_{\rm bb}$ (keV)& $2.39^{+0.33}_{-0.19}$ & $2.90^{+0.82}_{-0.44}$ & \nodata & \nodata & $0.90^{+0.20}_{-0.17}$ & $0.90^{+0.02}_{-0.01}$\\
 & $N_{\rm bb}$ ($10^{-2}$)& $2.6\pm0.1$ & $2.5^{+0.5}_{-0.2}$ & \nodata & \nodata & $2.3\pm0.1$ & $2.4\pm0.1$\\
 POWERLAW & $\Gamma$ & \nodata  & \nodata & $0.94^{+0.28}_{-0.12}$ & $0.68^{+0.29}_{-0.35}$ & $1.78\pm0.02$ & $1.80\pm0.02$\\
 & $N_{\rm pow}$ & \nodata & \nodata & $0.10\pm0.04$ & $0.05^{+0.05}_{-0.03}$ & $0.88\pm0.03$ & $0.87^{+0.03}_{-0.02}$\\
\tableline
 GAUSSIAN & $E_{\rm line}$ (keV) & $6.62^{+0.04}_{-0.13}$ & $6.48^{+0.09}_{-0.06}$ & $6.52^{+0.05}_{-0.07}$ & $6.43^{+0.05}_{-0.03}$ & $6.43^{+0.05}_{-0.03}$ & $6.40^{+0.02}$\\
 & $\sigma$ (keV) & $0.26^{+0.15}_{-0.06}$ & $0.61_{-0.11}$ & $0.41_{-0.03}$ & $0.61_{-0.03}$ & $0.41_{-0.01}$ & $0.61_{-0.02}$\\
 & $N_{\rm gau}$ ($10^{-3}$) & $1.9^{+1.1}_{-0.5}$ & $4.6^{+0.6}_{-1.4}$ & $3.1^{+0.3}_{-0.4}$ & $5.0\pm0.6$ & $3.1^{+0.3}_{-0.4}$ & $5.3\pm0.6$\\
 & EW (eV) & $48^{+11}_{-10}$ & $115^{+42}_{40}$ & $76^{+20}_{-16}$ & $123^{+75}_{-29}$ & $74^{+16}_{-14}$ & $128^{+41}_{-35}$\\
\tableline
$\chi^{2}/d.o.f.$ &  & 2052/1852 & 2051/1852 & 2072/1852 & 2056/1852 & 2184/1852 & 2148/1852
\enddata
\end{deluxetable*}

\section{Data Analysis}
\subsection{Continuum} \label{sec_con}

Continuum models for NS LMXBs can be degenerate, at least over narrow wavelength ranges, as we have here. 
They can be equally well explained by different continuum models consisting of a disk 
blackbody, a blackbody-like component (to fit the boundary layer emission) and a powerlaw or Comptonized
component (e.g., \citealt{Barret01,Lin07}). We test three different continuum models, which are a disk 
blackbody plus a blackbody (Model 1), a disk blackbody plus a powerlaw (Model 2), and a blackbody plus a
powerlaw (Model 3), in order to find 
out necessary spectral components required to interpret the spectrum of Serpens X-1 in 
this observation. As mentioned in section \ref{sec:datared}, we use the HEG $-1$ over 
the $2.0-8.0$ keV energy band. The Galactic absorption is modelled by TBABS in XSPEC, 
the accretion disk blackbody emission by DISKBB, and the thermal emission from the NS 
boundary layer by BBODY. 

An iron emission line has been clearly detected in the HEG $-1$ 
spectrum, and we start by modeling the feature with a Gaussian line (GAUSSIAN in XSPEC). In all 
our fits, we restrict the line energy to be in the range of $6.4-6.97$ keV, where neutral/ionized iron K$\alpha$ lines
can only appear. However, given the relatively narrow energy range of the continuum, we find that the width ($\sigma$) of the Gaussian tends to be large values (1 - 1.5 keV), and large normalizations giving equivalent widths (EWs) greater than 500 eV,
even when fixing the energy of the line to 6.4, 6.7 or 6.97 keV. Since this is not seen in neutron star LMXB spectra, we restrict the width of the Gaussian based on previous fits to Serpens~X-1. For instance, the broadest width that \citet{Ng10} get when fitting a Gaussian to {\it XMM} data of Serpens~X-1 is $\sigma=0.27^{+0.14}_{-0.11}$~keV. Fits to other archival spectra of Ser~X-1 tend to give slightly broader Gaussians than this.  For instance, averaging all the fits in \citet{Cackett12} we find an average Gaussian width of $\sigma = 0.61$ keV. Using these as a guide, we try spectral fits with  (a) $\sigma$ restricted to be less than the upper limit of 0.41~keV from Ng et al., and (b) $\sigma$ restricted to be less than 0.61~keV from \citet{Cackett12}.

\begin{figure}
\begin{center}
\includegraphics[scale=0.5]{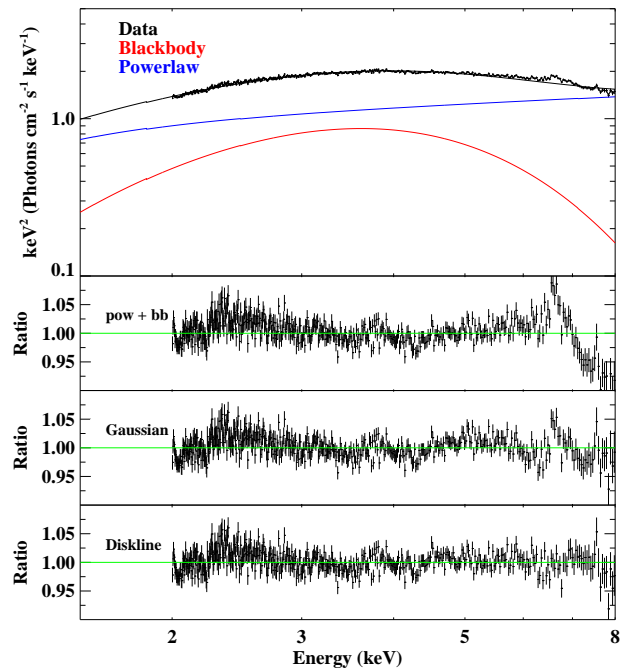}
\caption{The top panel shows the unfolded spectrum of Serpens X-1 with decomposed model components. 
We display the data/model ratios of the continuum model \texttt{tbabs*(bbody + powerlaw)}, Model 3 and Model 3a in lower panels.}
\label{eeuf}
\end{center}
\end{figure}

Including a broad Gaussian component in the iron line energy band significantly improves the fit (the smallest improvement was $\Delta \chi^2\sim150$ lower with three fewer d.o.f), confirming the clear detection of the broad iron emission line (also see differences between model residuals in Fig. \ref{eeuf}). In Table \ref{continuum} we show the fitting results of the three possible continuum models. Note that we also tried fitting with all three continuum components, but found that the power-law index became unconstrained in those fits. 

Models 1b \& 2a-b give comparable quality fits, however, all result in unphysical parameters. In Model 1b, when a larger $\sigma$ is allowed than Model 1a, we get an unusually high blackbody temperature (see Table \ref{continuum}). Furthermore, when replacing the Gaussian with a DISKLINE, we get an unphysical blackbody temperature ($>5$~keV), which is insensitive to the data (the peak of the blackbody is well outside of the HEG energy range).
A typical blackbody temperature seen in NS systems is generally below $\sim$ 2.5 keV\citep{Lin10,Lin12,Piraino12,DiSalvo15}.
In Model 2a-b the photon indices for the power-law are very hard ($\Gamma=0.94$), which is also unexpected
as photon indices of NS usually fall into the 1.5-3.0 range \citep{Cackett10,Cackett12,Egron13,DiSalvo15}.  
Therefore, while Model 3a-b do not give the best fit statistically, we use that model
for the continuum in the following analysis.  

It is interesting to note that both models 1a and 1b fit equally well, yet have quite different Gaussian widths. This can be viewed as evidence that the broad line is asymmetric, since a relativistic line with a narrow core and broad red wing can be well fit by two Gaussians - one reasonably narrow and one broad \citep{Cackett08,Cackett12}.  Fits with two Gaussians to {\it Suzaku} spectra of Ser~X-1 yield widths of 0.14 keV and 0.64 keV \citep{Cackett12}, comparable to the widths in models 1a and 1b.

\subsection{Iron Line} \label{sec_iron}

An iron emission line was clearly detected in the spectrum, implying that a reflection 
component is present in the system. The iron line of Serpens X-1 seems to extend for at 
least 1 keV, indicating possible signatures of relativistic effects similar to those seen in 
stellar-mass black holes and AGN. We test the nature of the line by fitting several different models.
If the line is broadened by relativistic effects, it should be better fitted by a relativistic line model rather
than a broad Gaussian. Given the high spectral resolution of the {\it Chandra} HEG, we also have
the opportunity to test whether there are any narrow line components that contribute to the line shape
that are otherwise unresolved with other detectors.

\begin{figure}
\begin{center}
\includegraphics[scale=0.5]{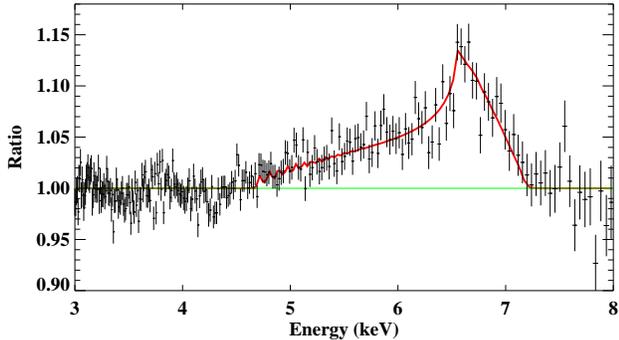}
\caption{The figure shows the Fe line profile of Serpens X-1. The spectrum is modeled as 
\texttt{tbabs(bbody + powerlaw + diskline)}, and the plot is produced by setting the normalization 
of DISKLINE to be 0. The plot is visually binned with ``septplot rebin" in XSPEC. The bin 
sizes through the Fe line are $\sim$ 0.042 keV.
The red line shows model. It can been seen that the DISKLINE model fits the asymmetric line
very well. }
\label{ironline}
\end{center}
\end{figure}

First, we tested the relativistic reflection model by fitting the Fe emission line using the 
DISKLINE \citep{Fabian89} model in XSPEC. We replaced the broad Gaussian component 
in Model 3a/3b with DISKLINE to build a new model (hereafter Model 4a). See Table \ref{comparison} for best-fitting values.
This model fits significantly better than a broad Gaussian, with an improvement of $\Delta \chi^2 = 133$ for two more
degrees of freedom. We find an inner disk radius of $7.7\pm0.1~R_{\rm G}$ (where $R_{\rm G} = GM/c^2$), and an
inclination of $24\pm1$ degrees.
We show the iron line profile and the best-fitting DISKLINE model in Fig. \ref{ironline}.  The DISKLINE component 
clearly fits the Fe line very well. Other relativistic line models such as LAOR and RELLINE were also tested, and all of
them fit the Fe line as well as DISKLINE does, with similar parameters.  Note that if we use Model 1a/1b for the continuum instead of Model 3a/3b, we get consistent DISKLINE parameters, and a large improvement in $\chi^2$ ($\Delta\chi^2\sim 72-73$), though because of the narrow energy range we get a high blackbody temperature which is unconstrained.

\begin{deluxetable*}{llrrrr}
\tabletypesize{\scriptsize}
\tablecaption{Best-fitting spectral parameters, testing different models for the Fe K emission line.  Model 4a tests a relativistic line, while models 4b tests narrow lines only and 4c tests a combination of a broad Gaussian
and narrow lines.  The relativistic line model is by far the best fit. Note that the definition of emissivity index
$\beta$ in the DISKLINE model is $\epsilon(r)=r^{\beta}$ (where $r$ is the disk radius) instead of 
$\epsilon(r)=r^{-q}$, and here we quote $q$, which is equal to $-\beta$.\label{comparison}}
\startdata
\tableline
Component &Parameter & Model 3b & Model 4a & Model 4b & Model 4c  \\
\tableline
TBABS & $N_{\rm H}$ ($10^{22}$ cm$^{-2}$)& (0.44) & (0.44) & (0.44) & (0.44) \\
BBODY & $kT_{\rm bb}$ (keV)& $0.90^{+0.02}_{-0.01}$ & $0.89^{+0.01}_{-0.02}$ & $0.89\pm0.01$ & $0.90\pm0.01$ \\
 & $N_{\rm bb}$ ($10^{-2}$) & $2.4\pm0.1$ & $2.4\pm0.1$ & $2.2\pm0.1$ & $2.3\pm0.1$ \\
POWERLAW & $\Gamma$& $1.80\pm0.02$ & $1.74\pm0.02$ & $1.75\pm0.02$ & $1.78\pm0.01$ \\
 & $N_{\rm pow}$ & $0.87^{+0.03}_{-0.02}$  & $0.81^{+0.03}_{-0.02}$ & $0.88^{+0.02}_{-0.03}$ & $0.88\pm0.01$ \\
\tableline
GAUSS 1 (broad) & $E_{\rm line1}$  & $6.40^{+0.02}$ & \nodata & \nodata & $6.4^{+0.03}$  \\
& $\sigma$ (keV) & $0.61_{-0.02}$ & \nodata & \nodata & $0.41_{-0.01}$ \\
 & $N_{\rm gau}$ ($10^{-3}$) & $5.3\pm0.6$ & \nodata & \nodata & $2.7\pm0.3$ \\
 & EW (eV) & $128^{+41}_{-35}$ & \nodata & \nodata & $64^{+16}_{-14}$ \\
GAUSS 2 (width=0) & $E_{\rm line2}$ & \nodata & \nodata & $6.57\pm0.01$ & $6.57\pm0.01$ \\
 & $N_{\rm gau}$ ($10^{-4}$) & \nodata & \nodata & $2.1^{+0.6}_{-0.4}$ & $1.4\pm0.5$ \\
 & EW (eV) & \nodata & \nodata & $5^{+2}_{-1}$ & $3^{+2}_{-1}$ \\
GAUSS 3 (width=0) & $E_{\rm line3}$ & \nodata & \nodata & $6.67\pm0.01$ & $6.67\pm0.01$  \\
 & $N_{\rm gau}$ ($10^{-4}$) & \nodata & \nodata  & $1.8^{+0.5}_{-0.4}$ & $1.2\pm0.5$ \\
 & EW (eV) & \nodata & \nodata & $5^{+1}_{-2}$ & $3\pm1$ \\
\tableline
DISKLINE & $E_{\rm line}$ (keV) & \nodata & $6.97_{-0.02}$ & \nodata & \nodata  \\
& $q$ & \nodata &$5.65^{+0.51}_{-0.43}$ & \nodata & \nodata  \\
& $R_{\rm in}$ ($RM/c^{2}$) & \nodata & $7.7\pm0.1$ & \nodata & \nodata  \\
& inclination & \nodata & $24\pm1$ & \nodata & \nodata  \\
& $N_{\rm diskline}$ ($10^{-3}$)  & \nodata & $5.8\pm0.5$ & \nodata & \nodata  \\
& $EW$ (eV)  &  \nodata & $149\pm15$ & \nodata & \nodata \\
\tableline
\tableline
$\chi^{2}/d.o.f.$ &  & 2148/1852 & 2015/1850 & 2286/1851 & 2149/1848 
\enddata
\end{deluxetable*}

\begin{figure}
\begin{center}
\includegraphics[scale=0.5]{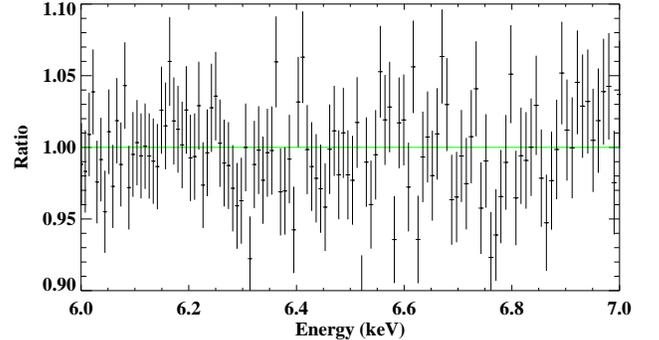}
\caption{The figure shows the 6.0-7.0 keV data/model ratio fitted using the DISKLINE model.
The data are unbinned with bin sizes of $\sim$ 0.008 keV.
It can be seen that no obvious emission or absorption features presenting in the figure.}
\label{ironline_2}
\end{center}
\end{figure}

Next, we tested if the line could be fitted by two narrow 
Gaussian components (Model 4b). The line width $\sigma$ of each narrow Gaussian 
component was set to be zero. Model 4b gives a significantly worse fit than Model 3b, indicating that
narrow lines alone cannot fit the data. We next tested for the presence of narrow lines in addition to
a broad component (Model 4c), which gave a comparable fit than Model 3b, but a significantly worse 
fit than the relativistic line ($\Delta \chi^2=134$ higher, with two fewer d.o.f.). A model including three 
narrow lines was tested but did not improve the fit. 
We also tried to set the line energies of the narrow Gaussian components to be those of the
Fe XXV (6.67 keV) and Fe XXVI (6.97 keV) lines, and again this did not yield a better fit.  
The equivalent widths of the narrow Gaussian lines in Model 4b and 4c are all small (EW $\sim3-5$ eV).
We also tested using narrow Gaussian components with physical upper limits of 
the line widths as well. Assuming a narrow line originates from outer part of the accretion 
disk and is broadened by thermal effects in a $\sim10^7$ K gas, the line width should be 
$\sigma\lesssim0.007$ keV.  This again made no improvement in the fit.  
From the series of tests, we conclude that there is no strong evidence of narrow emission lines.
We present detailed 6.0-7.0 keV data/model ratio in the lower panel of Fig. \ref{ironline_2}. It can
be clearly seen that Model 4a fits this band very well and no obvious emission or absorption features are present.

In conclusion, of all the models we tried to fit the Fe K line in Serpens~X-1, we find that a relativistic
line model fits significantly better than any others, which indicates that the Fe line profile is 
caused by relativistic broadening and no narrow line components are required to explain 
the spectrum. The result acts to validate the relativistic reflection scenario.

\subsection{Relativistic Reflection Model}

As the Fe emission line is best interpreted by the relativistic reflection scenario, we replace 
the DISKLINE component in Model 4a with a blurred broadband reflection model. The
reflection model self-consistently accounts for not only the Fe K line, but other 
lines and continuum emission expected due reflection, and broadening due to Comptonization
based on the ionization parameter.  The reflection model is then blurred by relativistic effects.

In NS systems, the accretion disk may be illuminated by the thermal emission coming from the 
boundary layer of the NS or by a power-law continuum, resulting in reflected emission 
\citep{Cackett10,DAi10}. The continuum of Serpens X-1 in this observation is dominated 
by the power-law component, and we use the REFLIONX model \citep{Ross05}, which 
calculates the broadband reflection spectrum from the accretion disk illuminated by a 
power-law continuum. The convolution model we use to account for relativistic effects is the 
KDBLUR kernel. We assumed the outer radius to be 400 $R_{\rm G}$. The iron abundance $A_{\rm Fe}$ was set to vary in 
the range between 1 and 4 times of solar value. The model fits the spectrum well, 
and we report the best-fitting parameters in Table \ref{reflection}. The values of the inner 
radius $R_{\rm in}$ and inclination angle $i$ we obtained are very similar to those given 
by Model 4a. The model using REFLIONX yielded a better fit than Model 4a
($\Delta \chi^2\sim 50$ lower with one fewer d.o.f.).

We also tested the XILLVER reflection model \citep{Garcia10,Garcia13}. While the fit using XILLVER
yielded a worse fit than that using the REFLIONX grid, but the best-fitting parameters are comparable to that of Model 4a. In
Table \ref{reflection} it can be seen that models using different reflection grids gave consistent
results. Parameters of KDBLUR obtained from both models are fairly similar, and both gave
a inner radius of $R_{\rm in}\sim7-8 R_{\rm G}$ and a low inclination angle ($i\sim30$).

\begin{table}
\caption{The table lists fitting results of the relativistic reflection models.
It can be seen that REFLIONX and XILLVER yield similar parameters.}
\label{reflection}
\centering
\begin{tabular}{@{}llrr}
\hline
Component & Parameter & REFLIONX & XILLVER\\
\hline
TBABS & $N_{\rm H}$  & (0.44) & (0.44)\\
BBODY & $kT_{\rm bb}$ (keV) & $0.86\pm0.01$ & $0.87\pm0.01$\\
 & $N_{\rm bb}$ ($10^{-2}$) & $2.7\pm0.1$ & $2.6\pm0.1$\\
POWERLAW & $\Gamma$ & $1.68\pm0.03$ & $1.75\pm0.03$\\
 & $N_{\rm pow}$ & $0.56\pm0.06$  & $0.73^{+0.02}_{-0.04}$\\
KDBLUR & $q$ & $5.1^{+1.8}_{-1.0}$ & $4.4^{+1.1}_{-0.7}$\\
 & $R_{\rm in}$ ($R_{\rm G}$) & $7.1^{+1.1}_{-0.6}$ & $8.4^{+1.1}_{-0.2}$\\
 & $i$ (deg) & $29\pm1$ & $33\pm1$\\
REFLECTION &$A_{\rm Fe}$ & $1.00^{+0.56}_{-0.11}$ & $1.52^{+1.47}_{-0.60}$\\
 & $\xi$ & $270^{+80}_{-30}$ & $130^{+70}_{-30}$\\
 & $N_{\rm ref}$ & $6.8^{+2.0}_{-2.3}\times10^{-5}$ & $0.19^{+0.64}_{-0.15}$\\
\hline
$\chi^{2}/d.o.f.$ & & 1968/1849 & 2012/1848 \\
\hline\\
\end{tabular}
\end{table}

\section{Discussion}

We analyzed a 300 ks {\it Chandra}/HEG observation of the NS LMXB Serpens~X-1.
We fit a number of models to the $2-8$ keV HEG $-1$ spectrum to examine the nature 
of the Fe emission line, and find that the origin of the line is best  explained by relativistically
broadened reflection.  Fitting broadband reflection models implies an inner radius of $\sim7-8$ $R_{\rm G}$ and a low inclination of 
$i\sim25^{\circ}-35^{\circ}$. In the following we discuss the robustness of the line 
parameters and compare our results with previous literature.

\subsection{Choice of continuum model}

It is difficult to constrain the continuum using a restricted $2-8$ keV energy band, and we 
choose the model with most reasonable fitting parameters (a blackbody and a powerlaw) 
to be the continuum in this work. In fact, Serpens X-1 has only been observed in the soft
state, and the powerlaw component is usually weak. To explain the spectrum of a NS in
the soft state, a continuum composed of a disk blackbody component contributed by the 
accretion disk and a blackbody component possibly caused by the thermal emission 
from the boundary layer (Model 1a/1b) is more likely. If replacing the Gaussian component with
the DISKLINE model in Model 1a/1b and re-fitting the spectrum, we still obtain DISKLINE
parameters similar to those of Model 4a ($E_{\rm line}=6.93\pm0.04$ keV; emissivity index 
= $-5.2^{+0.5}_{-0.3}$; $R_{\rm in} = 7.2\pm0.1$ $R_{\rm G}$; $i=26\pm1^{\circ}$). We
also conduct the same test on Model 2a/2b, and find that the choice of continuum
does not affect the parameters of the DISKLINE component.

Assuming the continuum of Serpens X-1 is soft and dominated by the thermal emission
from the boundary layer, the illuminating source is then the blackbody component. In order
to further test the broadband relativistic reflection model with a soft continuum (disk 
blackbody plus blackbody), we use the BBREFL grid (\citealt{Ballantyne04}; reflection
calculated assuming a blackbody component to illuminate the accretion disk) instead of 
REFLIONX to account for reflection. The model composed of a soft continuum and 
relativistic reflection (KDBLUR*BBREFL) gives the best-fitting inner radius 
$R_{\rm in}=8.1^{+0.7}_{-1.2}$ $R_{\rm G}$ and inclination angle $i=34^{+1}_{-3}$$^{\circ}$,
which are consistent with those obtained using a harder continuum with the
REFLIONX/XILLVER grids.

\subsection{No additional narrow line components}

In section \ref{sec_iron} we test various models to fit the Fe K line and show that the line
can simply be fitted by a relativistically broadened reflection component. Given the unique
spectral resolution of \emph{Chandra} HETGS, we have an opportunity to test for the presence
of any narrow lines, in addition to the broad line.  We find that including narrow lines in addition
to a broad Gaussian gives a worse fit than the relativistic line alone.  We include narrow Gaussian
components with reasonable line energies (6.4 keV, 6.67 keV and 6.97 keV) to Model 4a
and the broadband relativistic reflection models to examine the existence of narrow lines.
Narrow components at 6.4 keV and 6.67 keV do not improve the fit of 
Model 4a, and the narrow Gaussian line at 6.97 keV improves the fit marginally 
($\Delta \chi^2\sim 3$ lower than Model 4a). The narrow lines at 6.67 keV and 6.97 keV 
improve the fit of the REFLIONX model marginally ($\Delta \chi^2\sim 5-10$ lower than the 
REFLIONX model), but equivalent widths of these lines are low (EW $\sim 2$ eV). Furthermore, examination of Figure~\ref{ironline_2} shows no evidence for narrow lines in the residuals of the relativistic line fit. Hence, we conclude that there is no strong evidence of narrow components.

\begin{figure}
\begin{center}
\includegraphics[scale=0.5]{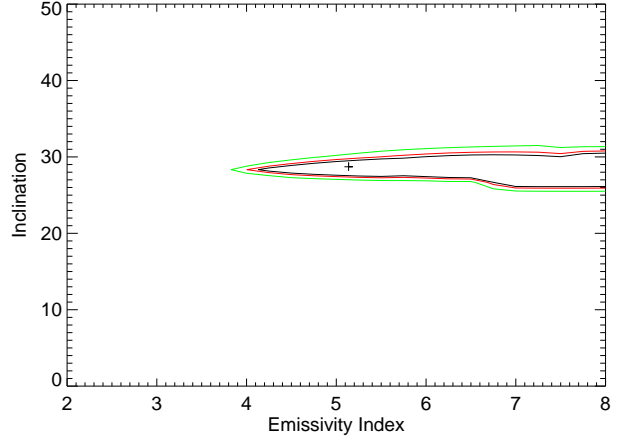}
\caption{The figure shows the contour plot of the emissivity index against the inclination angle,
which was calculated from the REFLIONX model. 
The contours are plotted at 67\% (black), 90\% (red) and 99\% (green) confidence levels. It can be seen
that there is no claer degeneracy between emissivity and inclination angle.
}
\label{contour}
\end{center}
\end{figure}

\subsection{Comparison with Previous Work}

\citet{Bhattacharyya07}, \citet{Cackett10} and \citet{Miller13} reported the existence of relativistic Fe K lines
in Serpens X-1 using \emph{XMM-Newton}, \emph{Suzaku} and \emph{NuSTAR} data, respectively.
In \citet{Bhattacharyya07} the continuum was modeled as an absorbed Comptonization (\texttt{compTT}) 
plus a disk blackbody, while in the later two pieces of work a full continuum (a disk blackbody,
a blackbody plus a powerlaw) was used.
\citet{Bhattacharyya07} used the LAOR model to fit the Fe K line, while \citet{Cackett10} and \citet{Miller13}
used the blurred BBREFL and REFLIONX models to account for relativistic reflection. \citet{Bhattacharyya07}
obtained a inclination angle of $i\sim40^{\circ}-50^{\circ}$, which is higher than the best-fitting values
of \citet{Cackett10} ($i\lesssim25^{\circ}$) and \citet{Miller13} ($i\lesssim20^{\circ}$) and this work ($i\sim30^{\circ}$).
\citet{Cackett10} also analyzed the same \emph{XMM-Newton} data set used in \citet{Bhattacharyya07}
and found that low inclination ($i\lesssim30^{\circ}$) is preferred. 

\citet{Cackett10} and \citet{Miller13} both obtained a low emissivity index of $q\sim2.3$, 
while a higher value of $q\sim4-5$ is required in this work.  We note that inclination and emissivity can be
degenerate \citep[see Fig. 8 in][]{Cackett10}, but the degeneracy was not found in this work (see Fig. \ref{contour}).
The emissivity index depends on the Fe K line profile, and it is likely that the line profile is slightly
different from those of previous observations because the better spectral resolution of \emph{Chandra}
grating data in this energy band.
Regardless, our fits suggest a small inner radius, consistent with previous findings.
In this work, the broad iron line profile implies an inner radius of $\sim7-8$ $R_{\rm G}$.
Previous analyses on \emph{XMM-Newton}, \emph{Suzaku} and \emph{NuSTAR} data suggest the inner radius
to be $\sim4-25$ $R_{\rm G}$, $\sim6-8$ $R_{\rm G}$, and $\sim6-8.5$ $R_{\rm G}$, respectively.
The \emph{XMM-Newton} observation was taken under timing mode with a short exposure time,
which causes some uncertainties in constraining parameters. The later \emph{Suzaku}, \emph{NuSTAR}
and current observations all give well-constrained, consistent measurements of the inner radius.
Although different continuum models were used to fit the spectra of Serpens X-1 observed 
with various instruments at different times/fluxes, previous and current analyses all indicate the source
to have low inclination angle and small inner radius. 

All previous observations of Serpens X-1 discussed above place the inner disk radius close to the ISCO.  
When comparing to the broader class of NS LMXBs, most other sources have an inner disk radius consistent 
with the ISCO, regardless of source luminosity \citep[for instance see Figure 7 of][for a comparison over two 
orders of magnitude in luminosity]{Cackett10}. Clear exceptions to this are SAX~J1808.4$-$3658 
\citep{Cackett09,papitto09}, IGR~J170480$-$2446 \citep{miller11}, and GRO~J1744$-$28 \citep{Degenaar14}, 
all of which are pulsars whose magnetic field must truncate the inner accretion flow. In low/hard states, accretion 
flow models suggest that the disk should recess, being replaced by a hot radiatively inefficient flow 
\citep[see e.g.,][for a review]{DGK07}. However, iron line studies of NS LXMBs in low states do not show large 
disk truncation. For instance, a {\it NuSTAR} observation of 4U 1608-52 at 1-2 per cent of the Eddington limit 
has a disk that extends close to the ISCO \citep{degenaar15}, while broadband {\it Suzaku} observations of 
4U~1705$-$44 at 3 per cent Eddington also shows the disk is not truncated at large radii \citep{DiSalvo15}. 
Thus, the majority of NS LMXBs, including Serpens X-1, show inner disk radii that are close to the ISCO, with 
no strong dependence on state or luminosity.

\section{Conclusion}

We analyze the latest long HETGS data of Serpens X-1 and examine the nature
of its Fe emission line. A thermal blackbody component possibly contributed by the 
boundary layer of the NS, and a power-law component provides a good fit to the continuum 
without unphysical parameters. By studying the Fe emission line in the spectrum, we 
find that the relativistic reflection scenario is much preferred, which is consistent with 
previous studies. \citet{Cackett10} analyzed \emph{Suzaku} data of Serpens X-1 and 
found relativistically-blurred iron emission lines. The recent \emph{NuSTAR} observation 
\citep{Miller13} confirms the presence of the relativistic iron line, together with the 
Compton hump. In this work, we construct several models to test the relativistic reflection
scenario and find that blurred reflection explains the Fe line profile significantly 
better than single/multiple Gaussian lines. Thanks to the remarkable resolving power of 
\emph{Chandra} HETGS, the grating spectrum is capable of detecting narrow
emission lines. In our analysis, no narrow line components are required, and must 
be weak if existent. 

The broad iron line profile implies a small inner radius, and we obtain an inner radius
of $\sim7-8 R_{\rm G}$. This sets an upper limit to the NS radius of $\sim15-17$ km
(assuming the mass of the NS is $\sim$ 1.4 $M_{\odot}$). 
Furthermore, a low inclination angle of $\sim25-35$ degrees is found, which is consistent with 
the previous measurements. We also find that the choice of continuum 
does not affect the values of the line-related parameters, which further confirms the 
robustness of the fitting results. We conclude that the Fe emission line observed in 
the X-ray spectrum of Serpens X-1 is broad and shaped by relativistic effects.

\acknowledgments
This work was greatly expedited thanks to the help of Jeremy Sanders in
optimizing the various convolution models. CYC and EMC gratefully 
acknowledge support provided by NASA through Chandra Award Number 
GO4-15041X issued by the Chandra X-ray Observatory Center, which is 
operated by the Smithsonian Astrophysical Observatory for and on behalf 
of NASA under contract NAS8-03060.

\bibliographystyle{apj}
\bibliography{serpensx1}

\appendix

\section{Discrepancy between the +1 and $-1$ Spectra} \label{sec_discrepancy}
\begin{figure}[h]
\begin{center}
\includegraphics[scale=0.5]{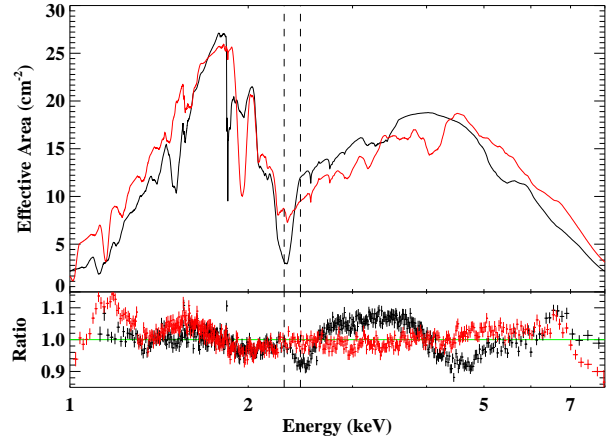}
\caption{We plot the HEG +1 (in black) and $-1$ (in red) detector effective areas in the upper panel.
The lower panel of the figure shows the Serpens X-1 HEG +1 (black data points) and 
$-1$ (red data points) data/model ratios. The continuum is modeled as 
\texttt{tbabs*(diskbb + bbody + powerlaw)}. The plot reveals the poor calibration of the HEG +1 spectrum.
The vertical dashed lines are intended to help guide the eye.}
\label{discrepancy}
\end{center}
\end{figure}

We aim at analyzing data with the maximum signal-to-noise ratio, so seek
to combine the HEG +1 and $-1$ spectra. The HEG $\pm1$ spectra are, however, 
discrepant and not suitable for combination. In Fig. \ref{discrepancy} we plot the 
detector effective areas and 
data/model ratio of the HEG $\pm1$ spectra (+1 in black and 
$-1$ in red data points) fitted by a simple continuum 
\texttt{tbabs*(diskbb + bbody + powerlaw)}. It can been seen
that the $\pm1$ spectra disagree in most of the HEG energy band (at the $5-10$\% level), though
both show similar Fe line profiles when the continuum is properly modeled.
The wiggles and emission features shown in the $-1$ spectrum below 2.0 keV cannot be modeled and match changes in the effective area, thus are likely due to calibration uncertainties. There are 
obvious deviations over the $2-5$ keV band between the $\pm1$ spectra. It seems
the HEG +1 spectrum suffers calibration problems, while the HEG $-1$ spectrum
does not show unexpected features in the $2-5$ keV energy band. 
The location of the zeroth order image on the chip determines at which energies the chip gaps lie. 
Here, a 1 arcmin y-offset was applied on the zeroth order in order to place it in a location to avoid any chip gaps near the Fe K band. This results in a chip gap at around 2.5 keV in the +1 spectrum. Use of such an offset is not widely reported previously,
which may explain the issues here.
The $\sim$2.5 keV drop in the +1 spectrum matches with a large change in the 
effective area due to a chip gap (as marked by dashed lines in Fig. \ref{discrepancy}). 
Uncertain calibration around this gap clearly leads to the residuals. 
There also seems to be excess emission around the $3-4$ keV.
This has been a known issue for HETGS data in CC mode, probably causing 
by improper order sorting table (OSIP) or charge transfer inefficiency (CTI) 
corrections$^1$. 

Different from the ``time event (TE)" mode, in CC mode every CCD suffers 
from time-dependent CTI. The CTI
correction relies on charge trap maps for each device to predict charge losses
and correct them. Nevertheless, in CC mode charges are clocked continuously 
and possibly leads to time-dependent charge trap maps, and each CCD suffers 
from this effect. Inappropriate CTI correction may cause events to fall out of the
OSIP. So far, alternate trap maps for CC mode are not found, but the 
\emph{Chandra} calibration team provided a few possible methods to solve this 
issue\footnote{http://cxc.harvard.edu/cal/Acis/Cal\_prods/ccmode/ccmode\_final\_doc02.pdf}.
One can primarily use HEG $-1$ and MEG +1 orders only, or apply a custom OSIP
to possibly fix the problem.

\begin{figure}
\begin{center}
\includegraphics[scale=0.5]{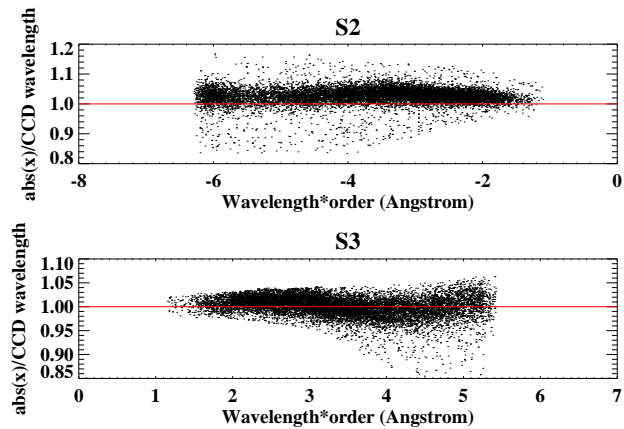}
\caption{The order-sorting region of chips S2 and S3. The x-axis is the wavelength times 
order, and the y-axis the absolute value of x-value over the CCD wavelength. The figure
shows the selected events during the order-sorting process. The mild curvature shown 
in S3 might be the cause of discrepancy between the HEG $\pm1$ spectra.}
\label{osip}
\end{center}
\end{figure}

We examined the order-sorting regions of each CCD by plotting the grating
wavelength (wavelength times order) against the wavelength over the CCD 
wavelength ($hc$/ENERGY). 
Only HEG first order data have been used, and we show the order-sorting regions 
of chips S2 and S3, where the $\sim$ $2-5$ keV $\pm1$ spectra were extracted from, 
in Fig. \ref{osip}. Ideally data points should be evenly distributed along the $y=1$ 
axis. Most events on chip S2 lie slightly above unity (and so as those on S1, S4 and 
S5), while the distribution of those on chip S3 shows mild curvatures, which might
be the reason that the +1 spectrum is not consistent with the $-1$ spectrum. 
One possible way to improve the
$\pm1$ spectral agreement is to modify the event file and bring the $y$-value in 
Fig. \ref{osip} close to unity. We first tried to correct the ENERGY column in the 
level=1.5 (after order-sorting) event file by dividing the column by the average 
$y$-value of each chip, and extract spectra from the corrected event file. We also
tried a more sophisticated method, which is modifying the ENERGY column node 
by node, i.e., applying a spline fit to make every event lies on $y=1$. However, neither of the 
methods made the spectra noticeably better. We then tried to modify the
level=1 (before order-sorting) event file and re-run \texttt{tg\_resolve\_events} to 
process order-sorting on the corrected event file, and extracted a new spectrum
from the new event file after order-sorting. Modifying the level=1 event file does give 
different spectra, but it seems it causes problems to order-sorting and the spectra 
are clearly not corrected properly.

Another method to tackle this issue is to widen the order-sorting window. When data
are taken under the CC mode, the Y position of an event is not known, and the order 
selection can be tricky. In some cases the 
spectra can be smoothed out by including events fell out of the OSIP. When running
\texttt{tg\_resolve\_events}, by setting the script to disable the original OSIP file and 
indicating numbers of parameters ``osort\_hi" and  ``osort\_lo", one can customize the size 
of the order-sorting window. We set both osort\_hi and osort\_lo to be 0.2 (including events
fell in the regime $0.8<y<1.2$ in Fig. \ref{osip}) to widen the event order-sorting window.
We find that this does not help eliminate discrepancies between the
$\pm1$ spectra, but on the contrary, makes the issue worse. By widening the order-sorting
window, more events are selected and hence spectra with higher fluxes are created.
The 3-4 keV excess emission in the +1 spectrum turned to be larger than the original 
spectrum before correction. In this case, widening the order-sorting window does not 
effectively solve the problem.

We look for a correction that would remove the 3-4 keV excess from the HEG +1 spectrum.
As widening the order-sorting would increase the flux of this energy band, narrowing the
window might give a correction that we need.  
We find that setting osort\_hi=0.2 and osort\_lo=0.04 (including events
fell in the regime $0.96<y<1.2$) improves the agreement of the $\pm1$
spectra. The corrected +1 spectrum does not completely match the $-1$ spectrum, but the 
flux level between $2-5$ keV is much more similar than the spectrum before correction.
Although unfortunately it is still not good enough to combine the $\pm1$ spectra and 
achieve maximum signal-to-noise ratio, we find that applying a SMEDGE component in 
XSPEC with negative optical depth to the +1 spectrum when fitting significantly improves 
data agreement over the $2-5$ keV energy band.
The SMEDGE component can mimic the sharp feature caused by the change in 
effective area and reduce the residuals. Yet it seems that different continuum models are
required to fit the +1 and $-1$ spectra, thus we only present results of the $-1$ spectrum in the 
paper.  Nevertheless, fitting the +1 spectrum including the SMEDGE and allowing for a different
continuum model than the $-1$ spectrum, results in the same conclusions and consistent parameters
for the relativistic iron line.

\end{document}